\documentclass[%
superscriptaddress,
showpacs,
 amsmath,amssymb,
prb,
preprint,
floatfix,
endfloats*
]{revtex4-1}
\usepackage{graphicx}
\usepackage{bm}
\usepackage[breaklinks=true,urlcolor=blue,bookmarks=true]{hyperref} 
\usepackage{color}

\newcommand{\X}{\mathrm{X}}

\begin{document}

\title{Charge-Tunable Indium Gallium Nitride Quantum Dots}

\author{Lei Zhang}
\affiliation{Physics Department, University of Michigan, 450 Church Street, Ann Arbor, MI 48109-1040, USA}
\author{Chu-Hsiang Teng}
\affiliation{EECS Department, University of Michigan, 1301 Beal Avenue, Ann Arbor, MI 48109-2122, USA}
\author{Pei-Cheng Ku}
\affiliation{EECS Department, University of Michigan, 1301 Beal Avenue, Ann Arbor, MI 48109-2122, USA}
\author{Hui Deng}
\affiliation{Physics Department, University of Michigan, 450 Church Street, Ann Arbor, MI 48109-1040, USA}

\begin{abstract}
III-Nitride quantum dots have emerged as a new chip-scale system for quantum information science, which combines electrical and optical interfaces on a semiconductor chip that is compatible with non-cryogenic operating temperatures. Yet most work has been limited to optical excitations. To enable single-spin based quantum optical and quantum information research, we demonstrate here quantized charging in optically active, site-controlled III-Nitride quantum dots. Single-electron charging was confirmed by the voltage dependence of the energy, dipole moment, fine structures and polarization properties of the exciton states in the quantum dots. The fundamental energy structures of the quantum dots were identified, including neutral and charged excitons, fine structures of excitons, and A and B excitons. The results lay the ground for coherent control of single charges in III-Nitride QDs, opening a door to III-Nitride based spintronics and spin-qubit quantum information processing.
\end{abstract}

\maketitle

Charge-tunable semiconductor quantum dots have been an important systems in quantum information science and technology \cite{warburton_single_2013, kloeffel_prospects_2013}. This is because they uniquely provide a chip-scale platform of long-lived single spins \cite{warburton_optical_2000} with an ultrafast optical interface via the exciton transitions \cite{gerardot_optical_2005, atature_quantum-dot_2006, koppens_driven_2006, xu_fast_2007, nowack_coherent_2007, press_complete_2008}.
However, most work has been performed on self-assembled III-Arsenide QDs, which are intrinsically limited to cryogenic operating temperatures. The random positioning of these QDs also limits their scalability.

Recently, wide-bandgap III-Nitride QDs have emerged as an alternative semiconductor system with the potential for higher operating temperatures, better scalability and longer spin relaxation times. Single photon emission has been reported at high temperatures \cite{kako_gallium_2006, holmes_room-temperature_2014}, from site controlled structures \cite{hsu_single_2011, zhang_single_2013, holmes_room-temperature_2014}, and with pre-defined polarizations \cite{lundskog_direct_2014,teng_elliptical_2015}. Exciton Rabi oscillation at low temperatures was demonstrated recently with potential applications as an exciton qubit \cite{holmes_measurement_2013}.
Coulomb blockade was also observed in electrical devices via transport measurements \cite{songmuang_quantum_2010}.
However, no work has been reported to date on their potential to host optically-coupled spin-qubits for quantum information processing, a prerequisite of which is charge-tuning of optically active QDs.

In this work, we demonstrate quantized charge-tuning of an optically stable site-controlled InGaN/GaN QDs, confirmed via discrete changes in the emission energy, permanent dipole moment, and spin-orbit interactions in the excitonic excitations. Furthermore, the charge-control of the QDs, together with the high optical quality, allowed us to unambiguously identify the basic energy structures of the QDs, including simultaneous identification of the exciton fine-structures, the exciton and trion splitting, and the A- and B-exciton splitting. Quantitative knowledge of the exciton fine-structures, in particular, is crucial for entanglement generation using QDs. These results are enabling steps toward coherent control of single charges in III-Nitride QDs and III-Nitride based spintronics and spin-qubit quantum information processing.

\begin{figure*}
\includegraphics[width = 0.9\textwidth]{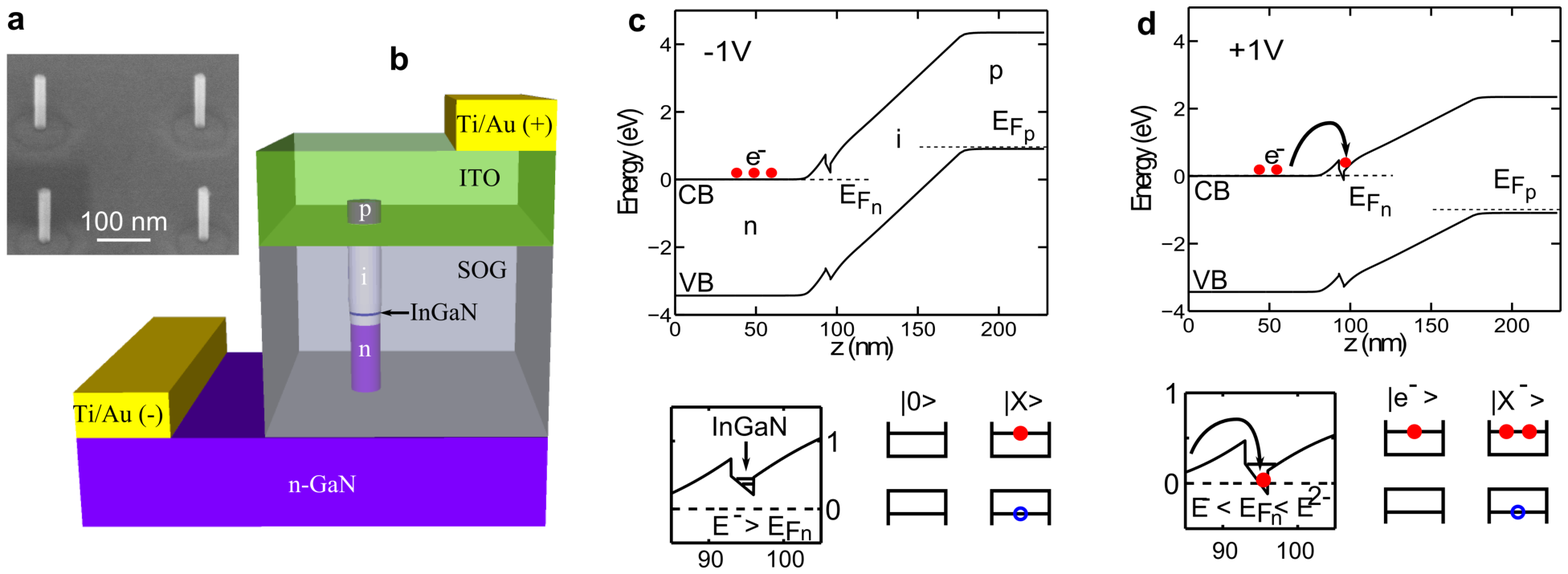}
  \caption{\label{fig:sample} The sample structure and operating principle. \textbf{a}, 45$^\circ$-angle SEM image of an array of four nanowires before the fabrication of electrical contacts. Single QDs are located in the middle of the nanowires. 
  \textbf{b}, A schematic plot of a single QD diode. The 3-nm thick In$_{0.15}$Ga$_{0.85}$N QD is sandwiched by a 50-nm thick Mg-doped GaN (p-GaN) and 80-nm thick i-GaN layer at the top, and a 10-nm thick i-GaN and 90-nm thick Si-doped GaN (n-GaN) layer at the bottom.
  \textbf{c},\textbf{d}, Calculated conduction band (CB) and valence band (VB) profiles along the growth direction (z-axis) at $V_\mathrm{bias} = -1$~V and 1~V, respectively. The lower-left panels are zoomed-in view of the CB near the InGaN region, illustrating single electron ($e^{-}$) charging as the electron quasi-Fermi level ($E_\mathrm{Fn}$) moves above the first charging energy $E^{-}$. The lower-right panels illustrate the ground and excitonic states of the neutral (c) and charged (d) QDs.}
\end{figure*}

To enable both electrical and optical controls efficiently and reproducibly, we used InGaN QDs created by a top-down approach \cite{lee_fabrication_2012,zhang_single_2013} in a compact nano-wire structure without a wetting layer, as shown in Fig.~\ref{fig:sample}. For electrical control, these QDs allow a similar planar junction as widely used III-N light-emitting diodes with low serial resistance and no wetting-layer induced leakage current. For optical control, fast single photon emission has been demonstrated free from background luminescence typically associated with wetting-layers \cite{zhang_single_2013}. Moreover, with the position and size of the QDs controlled to within a few nano-meters \cite{lee_fabrication_2012,zhang_single_2013}, each individual QD can be identified, isolated and repeatedly accessed.

Single electron charging was achieved based on Coulomb blockade when the electron quasi-Fermi level $E_\mathrm{Fn}$ was tuned across quantized conduction-band states in a QD. To illustrate this process, we simulated the voltage-controlled carrier concentrations and the Fermi energies of the conduction band (CB) and valence band (VB), $E_\mathrm{Fn}$ and $E_\mathrm{Fp}$ (Appendix~\ref{appedix:simulation}). We solve first the three-dimensional (3D) strain distribution and then self-consistently the band structure and drift-diffusion equations \cite{sacconi_optoelectronic_2012}.
As shown in Fig.~\ref{fig:sample}(c) and \ref{fig:sample}(d), at a negative or very small positive bias voltages $V_\mathrm{bias}$, $E_\mathrm{Fn}$ lies below the single-electron charging energy $E^{-}$ and the QD is neutral.
When $V_\mathrm{bias}$ is increased such that $E_\mathrm{Fn}>E^{-}$, an electron can tunnel from n-GaN into the QD. The presence of an electron inside the QD prevents additional electrons from entering the QD due to Coulomb blockade. Hence the QD will stay singly charged until $V_\mathrm{bias}$ increases further to overcome the Coulomb blockade, reaching above the two-electron charging energy $E^{2-}$. Corresponding to the different charge occupations of the QDs, neutral or charged excitons are created under optical excitations (lower-right panels in Fig.~\ref{fig:sample}(c), \ref{fig:sample}(d)). They feature different energies, permanent dipole moments and degeneracies, providing experimental signatures for the charge occupation in the QD.

To create the electrically-controlled InGaN QDs, single InGaN/GaN quantum well was patterned via electron-beam lithography followed by reactive-ion etching \cite{lee_fabrication_2012} and selective wet-etch in buffered KOH solution \cite{li_optical_2011}. To create the electrical contacts, the nanowires were first planerized by a 580~nm thick spin-on-glass, which was then etched back using CF$_4$-CHF$_3$ plasma to expose 30~nm of p-GaN layer. A semi-transparent indium-tin-oxide p-contact was then deposited by DC sputtering, patterned using HCl and annealed at 400~$^\circ$C for 5 minutes in forming gas. Both n- and p-contacts were metalized using Ti/Au (25~nm/500~nm) pads. In this structure, the positions and sizes of the QDs are precisely controlled to within a few nanometers \cite{lee_fabrication_2012,zhang_single_2013}. Hence each individual QD can be identified, isolated and repeatedly accessed. There was no wetting layer, and thus no leakage current or background luminescence was measured.

To characterize the optical properties of the QDs, we used a confocal micro-photoluminesence ($\mu$-PL) setup to isolate PL from single QDs with 0.8~$\mu$m spatial resolution and 170~$\mu$eV spectral resolution at $3$~eV \cite{zhang_single_2013}. Polarization-dependent PL measurements were performed using a rotating half-wave plate and a fixed polarizer. The sample was kept at 10 K in a He-flow cryostat with voltage feed-through ports and was excited by a 370 nm femtosecond pulsed laser. To allow $\mu$-PL study of single QDs, an array with a QD-QD spacing of 20-$\mu$m was used.

\begin{figure*}
	\includegraphics[width = 0.9\textwidth]{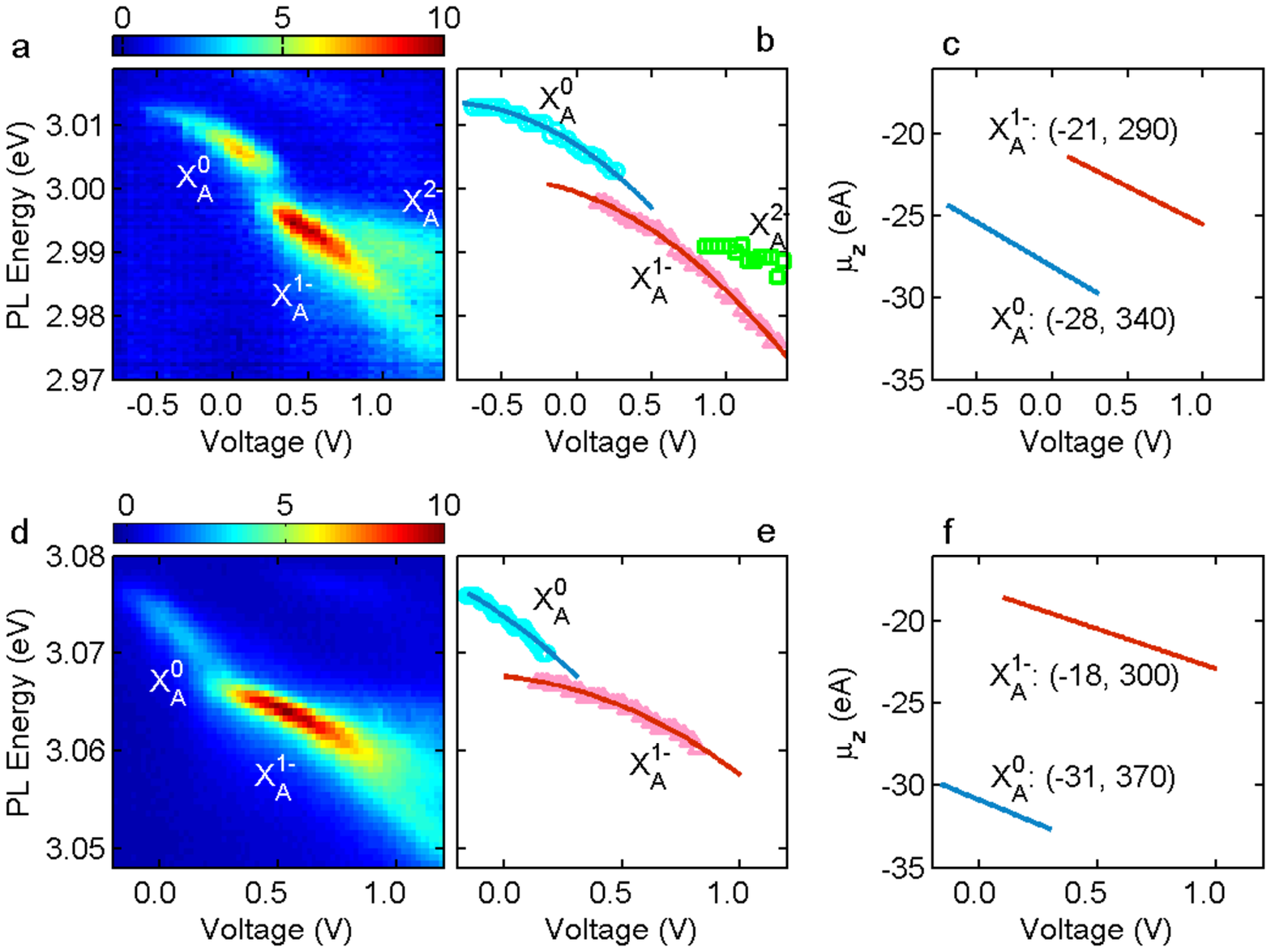}
	\caption{\label{fig:qd1_vs_qd2} Voltage-dependent PL spectra, peak energies and static dipole moments of the excitons from two QDs, QD1 (\textbf{a}-\textbf{c}) and QD2 (\textbf{d}-\textbf{f}).
\textbf{a},\textbf{d}, The $\mu-$PL spectra vs. the bias voltage $V_\mathrm{bias}$, showing discrete bands corresponding to neutral and charged A- and B- excitons. \textbf{b},\textbf{e}, The emission peaks extracted from \textbf{a} and \textbf{d}, respectively. The solid curves are parabolic fits using equation~(\ref{eq:deltaE}), which yield the dipole moment $\mu_\mathrm{z0}$ at $V_\mathrm{bias} = 0$ and polarizability $\alpha_\mathrm{z}$. \textbf{c},\textbf{f}, $\mu_\mathrm{z}$ vs. $V_\mathrm{bias}$ obtained using equation~(\ref{eq:muz}) and the fitted values of $\mu_\mathrm{z0}$ and $\alpha_\mathrm{z})$. For each state, the $\mu_\mathrm{z0}$ in e$\mathrm{\AA}$ and $\alpha_\mathrm{z}$ in meV/(MV/cm)$^2$ are labeled in the figure in the parentheses as $(\mu_\mathrm{z0}, \alpha_\mathrm{z})$.}
\end{figure*}

Figure~\ref{fig:qd1_vs_qd2} shows the photoluminescence (PL) spectra of A-excitons from two QDs, QD1 and QD2, versus the bias voltage $V_\mathrm{bias}$. The voltages are corrected from the laser induced photo-voltage \cite{seidl_absorption_2005} as described in Appendix~\ref{appedix:photovoltage}. Discrete emission bands were observed, corresponding to different charging states of the QD. No emission was observed at large negative bias. With $V_\mathrm{bias}$ increased to above $-0.6$~V, emission from neutral exciton $\X_\mathrm{A}^0$ was observed from QDs that were initially empty. 
With $V_\mathrm{bias}$ increased to $V_1\sim 0.25$~V, the emission peak red-shifted suddenly, showing that one electron tunneled into the QD. The energy $E_\mathrm{A}^{1-}$ of the singly charged-exciton $\X_\mathrm{A}^{1-}$ was lower compared to $\X_\mathrm{A}^{0}$, suggesting that the attractive Coulomb interactions between two electrons and one hole in the QD exceeded the repulsive Coulomb interaction between the two electrons. With $V_\mathrm{bias}> V_2\sim 0.8$~V, an additional peak $\X_\mathrm{A}^{2-}$ appeared at an energy higher than $E_\mathrm{A}^{1-}$, corresponding to the tunneling of an additional electron into the QD, which introduced strong repulsive Coulomb interaction among the electrons. The co-existence of multiple emission peaks at $V_\mathrm{bias}\sim V_1$ and $V_2$ was due to charges co-tunneling between the GaN barrier and the QDs \cite{baier_optical_2001}. 

Due to the tight confinement in a QD, even single charge occupation can change significantly the internal electric field and thus the exciton static dipole moment and the exciton energy, which we can obtain from the voltage dependence of the exciton energy. 
As shown in Fig.~\ref{fig:qd1_vs_qd2}(b) and \ref{fig:qd1_vs_qd2}(e), the exciton resonances red-shifted super-linearly with an increasing $V_\mathrm{bias}$, suggesting enhanced quantum confined Stark effect with changing internal electric field. 
The total electric field $F_\mathrm{tot}$ in the QD consists of two components: a fixed internal polarization field $F_\mathrm{pol}$ and a voltage-dependent external field $F_\mathrm{pin}$ across the p-i-n junction.
$F_\mathrm{pol}$ originates from the piezoelectric and spontaneous polarization charges at the InGaN/GaN interfaces; it is determined by the composition and structure of the QD and is independent from $V_\mathrm{bias}$. $F_\mathrm{pin}$ is due to the separation of the electrons and holes across the p-i-n junction; it varies linearly with $V_\mathrm{bias}$ as $F_\mathrm{pin}=F_\mathrm{pin0} - \eta V_\mathrm{bias}$.
From numerical simulations, we found $F_\mathrm{pol}=-1.6$~MV/cm at the center of the QD, pointing from p-GaN to n-GaN as denoted by the negative sign, $F_\mathrm{pin0}=0.55$~MV/cm, and $\eta=0.16\times 10^6$/cm (Appendix~\ref{appedix:simulation}). Within the range of $V_\mathrm{bias}$ measured (Fig.~\ref{fig:qd1_vs_qd2}), $|F_\mathrm{pol}|\gg|F_\mathrm{pin}|$. Hence the exciton's static dipole moment $\mu_\mathrm{z}$ points opposite to $F_\mathrm{pin}$ and increases linearly with increasing $V_\mathrm{bias}$ according to \cite{empedocles_quantum-confined_1997}:
\begin{equation}
\mu_\mathrm{z}(V_\mathrm{bias}) - \mu_\mathrm{z0} = \alpha_\mathrm{z} (F_\mathrm{pin}-F_\mathrm{pin0})= - \alpha_\mathrm{z} \eta V_\mathrm{bias}.
\label{eq:muz}
\end{equation}
Here $\alpha_\mathrm{z}$ is the polarizability in the z-direction, and $\mu_\mathrm{z0}=\mu_\mathrm{z}(0)$ is the permanent dipole moment at zero bias. Correspondingly, the exciton energy shifts by
\begin{align}
\Delta E (V_\mathrm{bias}) & = -\mu_\mathrm{z}F_\mathrm{pin} + \mu_\mathrm{z0}F_\mathrm{pin0} \nonumber \\
& = (\mu_\mathrm{z0} + \alpha_\mathrm{z}F_\mathrm{pin0})\eta V_\mathrm{bias} - \alpha_\mathrm{z} \eta^2 V_\mathrm{bias}^2.
\label{eq:deltaE}
\end{align}
Fitting the measured exciton energy $E$ vs. $V_\mathrm{bias}$ from Fig.~\ref{fig:qd1_vs_qd2}(b) and \ref{fig:qd1_vs_qd2}(e) with equation~(\ref{eq:deltaE}), we obtained $\mu_\mathrm{z0}$ and $\alpha_\mathrm{z}$ for each excitonic state as labeled by the two numbers $(\mu_\mathrm{z0},\alpha_\mathrm{z})$ in the parentheses in Fig.~\ref{fig:qd1_vs_qd2}(c) and \ref{fig:qd1_vs_qd2}(f).
$\mu_\mathrm{z0}$ were negative, suggesting that the electrons and holes were concentrated at the top and bottom of the InGaN/GaN interfaces, respectively, consistent with the numerically calculated ground state electron and hole envelope functions (Appendix~\ref{appedix:simulation}). The magnitudes of $\mu_\mathrm{z0}$ were around $30$~$\mathrm{e\AA}$, consistent with the thickness of the InGaN layer of $30$~$\mathrm{\AA}$. The polarizability $\alpha_\mathrm{z}$ have values of $\sim 300$~meV/(MV/cm)$^2$, or $4\times 10^4$~$\mathrm{\AA}^3$ in cgs units, comparable to values reported from other semiconductor QDs \cite{wang_exciton_2006, jarjour_control_2007}.

With the values of $\mu_\mathrm{z0}$ and $\alpha_\mathrm{z}$ measured above, we obtained $\mu_\mathrm{z}$ vs. $V_\mathrm{bias}$ as plotted in Fig.~\ref{fig:qd1_vs_qd2}(c),\ref{fig:qd1_vs_qd2}(f). At $V_\mathrm{bias}=V_1$, accompanying the jump of the exciton emission energies from $\X_\mathrm{A}^0$ to $\X_\mathrm{A}^{1-}$, the magnitude $|\mu_\mathrm{z}|$ dropped suddenly by about $10$~e$\mathrm{\AA}$. This was also manifested in Fig.~\ref{fig:qd1_vs_qd2}(a), \ref{fig:qd1_vs_qd2}(b), \ref{fig:qd1_vs_qd2}(d), \ref{fig:qd1_vs_qd2}(e), where the slope of the exciton energy vs. the voltages changed suddenly from the $\X_\mathrm{A}^0$ band to $\X_\mathrm{A}^{1-}$ band. The change in $\mu_\mathrm{z0}$ suggests that, when an electron tunnels into the QD, it effectively screens the internal electric field of the QD and brings the electron and hole envelope functions of an charged exciton closer to the center of the QD through Coulomb interactions.

\begin{figure*}
\includegraphics[width =0.9\textwidth]{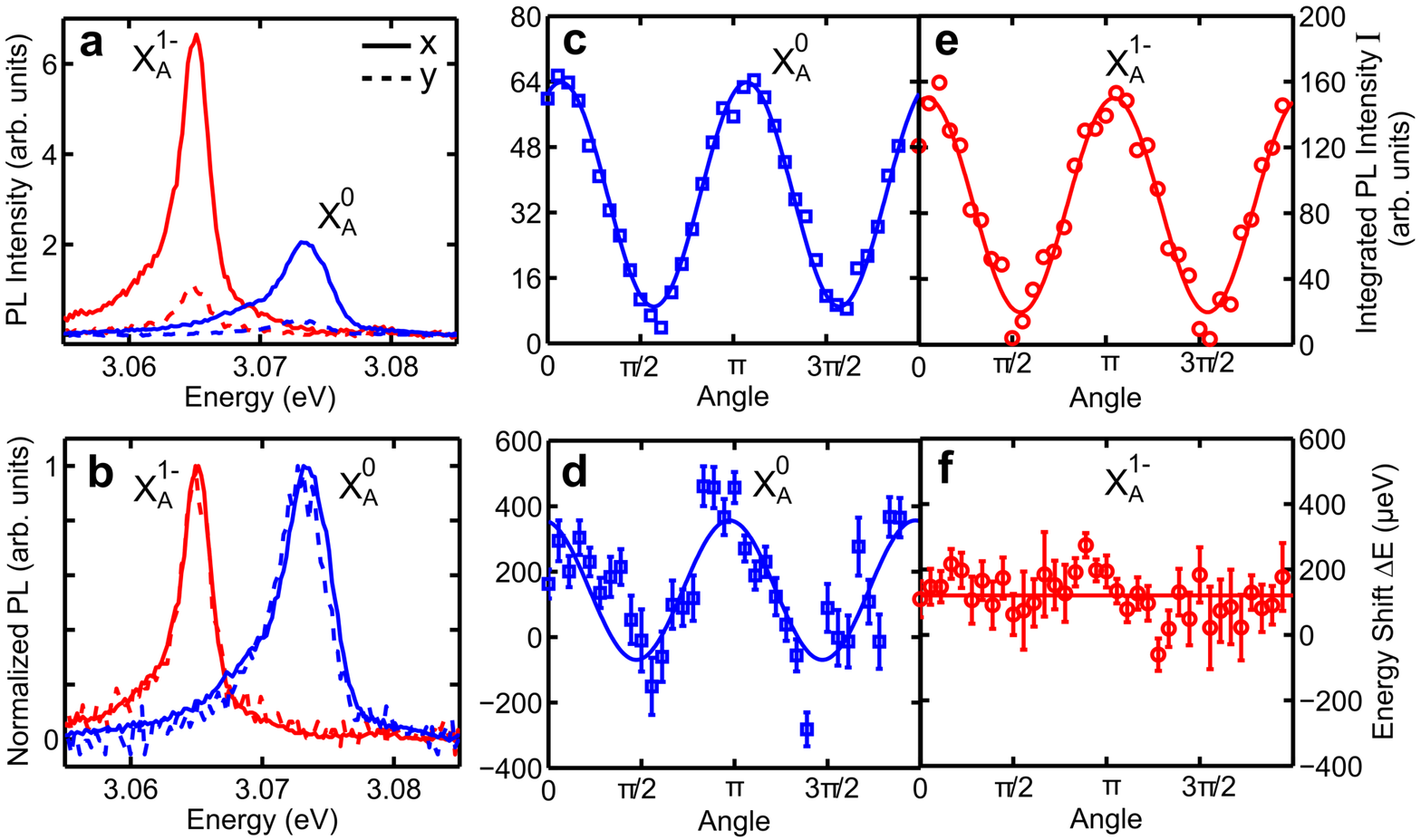}
\caption{\label{fig:AX_vs_CAX} Fine structures of the excitons by polarization-dependent PL of QD2. \textbf{a}, PL spectra of $\X_\mathrm{A}^0$ ($V_\mathrm{bias}=0$~V) and $\X_\mathrm{A}^{1-}$ ($V_\mathrm{bias}=0.5$~V) with x-polarization (solid lines) and y-polarization (dashed lines). The x direction was chosen as the direction of maximal $\X_\mathrm{A}^{0}$ intensity. \textbf{b}, The normalized spectra of \textbf{a} for easier visualization of any spectral shift between the two polarizations.  \textbf{c},\textbf{d}, The integrated intensities $I$ and energy shifts $\Delta E$ of $\X_\mathrm{A}^{0}$ vs. the polarization angles $\theta$. \textbf{e},\textbf{f}, The $I$ and $\Delta E$ of $\X_\mathrm{A}^{1-}$ vs. $\theta$. The error bars in $\Delta E$ as shown in \textbf{d} and \textbf{f} are from Gaussian fits of the PL peaks. The solid lines in \textbf{c}-\textbf{e} are independent fits using $a\cos^2(\theta - \theta_0) + b\sin^2(\theta - \theta_0)$, with $a$, $b$ and $\theta_0$ as the fitting parameters. The solid line in \textbf{f} marks the average value.}
\end{figure*}

To further confirm single electron charging of the QD, we measured the energy degeneracies of the exciton states by polarization-resolved PL.
As shown in Fig~\ref{fig:AX_vs_CAX}, the PL of $\X_\mathrm{A}^0$ consisted of two orthogonally polarized peaks with slightly different energies, while no energy difference could be measured between the two polarizations of $\X_\mathrm{A}^{-1}$ emission. These features were measured more accurately by the integrated PL intensity $I$ and the shift of the exciton energy $\Delta E$ vs. the polarization angle $\theta$. Both $\X_\mathrm{A}^{0}$ and $\X_\mathrm{A}^{-1}$ exhibited a high degree of linear polarization as shown by the sinusoidal oscillation of $I$ vs. $\theta$ (Fig.~\ref{fig:AX_vs_CAX}(c), \ref{fig:AX_vs_CAX}(e)). For $\X_\mathrm{A}^0$, the same oscillation was also measured for $\Delta E$ vs. $\theta$, clearly revealing an energy shift between the two polarizations (Fig.~\ref{fig:AX_vs_CAX}(d)). Fitting the data with a sinusoidal function yields the energy shift to be $370\pm 120$~$\mu$eV, in agreeement with previous reported values for confined InGaN/GaN nanostructures \cite{amloy_polarization-resolved_2011}. In contrast, no significant energy shift was observed for the charged exciton at different polarization angles (Fig.~\ref{fig:AX_vs_CAX}(f)).

\begin{figure*}
\includegraphics[width = 0.9\textwidth]{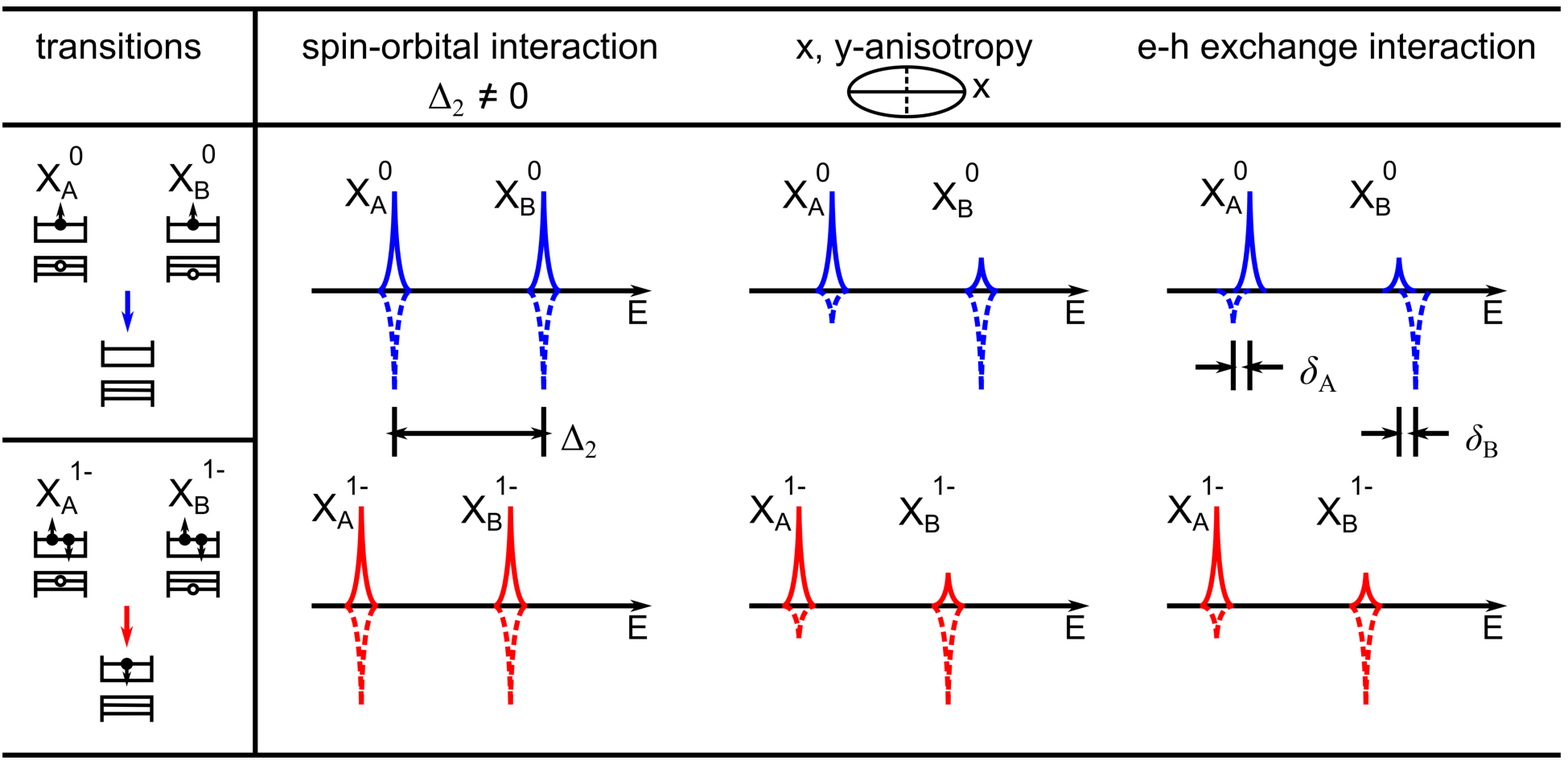}
\caption{\label{fig:fss-theory} Illustration of spectral splitting and polarization due to spin-orbital interactions, QD anisotropy and electron-hole exchange interactions in III-Nitride QDs \cite{bardoux_polarized_2008}. The blue and red spectral peaks denote neutral and charged exciton emissions, respectively. The solid, up-ward (dashed, down-ward) peaks denote the x-polarized (y-polarized) exciton emission, and the height of the peak indicates the relative oscillator strength or the emission intensity.}
\end{figure*}

To understand the polarization resolved measurement above, the energy levels and polarization states of neutral and charged QDs are illustrated in Fig.~\ref{fig:fss-theory}.
The lowest optical excitations in III-Nitride QDs correspond to the pairing of an s-orbital electron from the conduction band and a p-orbital hole from the A or B valence band, each with a two-fold degeneracy. The splitting between the A and B valence bands is mainly due to the spin-orbital interaction \cite{chuang_kp_1996}. Due to fabrication uncertainties, the QDs have slightly elliptical shapes and thus anisotropic strain distributions. Hence the two-fold degenerate A- and B- excitons have nearly linear polarizations, parallel and perpendicular to the elongated direction (x) of the QD, respectively \cite{winkelnkemper_polarized_2007, winkelnkemper_polarized_2008}. In a neutral QD, the two-fold degeneracy is further lifted by the electron-hole Coulomb exchange interactions, resulting in the fine structure splitting (FSS) of the neutral excitons \cite{bardoux_polarized_2008}
as was observed for $\X_\mathrm{A}^{0}$ in Fig.~\ref{fig:AX_vs_CAX}(d). In contrast, in a singly negatively charged QD, trions created by optical excitations have two conduction-band electrons forming a singlet with a total spin of zero. As a result, the electron-hole Coulomb exchange interaction vanishes and FSS becomes absent. This explains the suppressed energy shift observed for $\X_\mathrm{A}^{-1}$ in Fig.~\ref{fig:AX_vs_CAX}(f). 

\begin{figure*}
\includegraphics[width = 0.9\textwidth]{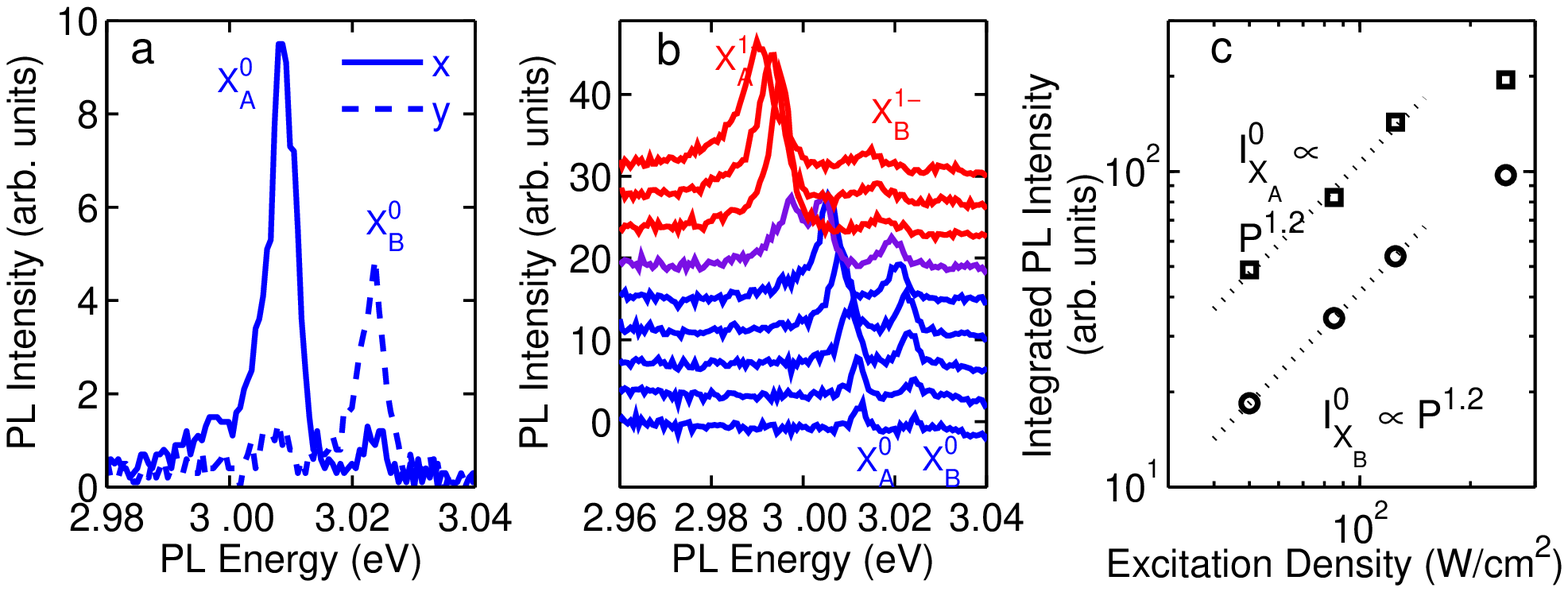}
\caption{\label{fig:AX_vs_BX} A-B exciton splitting in QD1. \textbf{a}, PL spectra of A- and B-excitons in neutral QD1 at $V_\mathrm{bias} = -0.04$~V for the x-polarization (solid line) and y-polarizations (dashed line). \textbf{b}, PL spectra of A- and B-excitons in neutral (red) and charged (blue) QD1 at $V_\mathrm{bias}$ from -0.58~V (bottom) to 0.70~V (top). \textbf{c}, The integrated intensity of $\X_\mathrm{A}^0$ and $\X_\mathrm{B}^0$ at $V_\mathrm{bias} = -0.04$~V vs. the excitation densities.}
\end{figure*}

By changing the QD's charge occupation, we were also able to distinguish the FSS from the A- and B-excition splitting. It has been difficult to distinguish them using neutral QDs because, similar to the fine structures, the A- and B-excitons have different intensities and orthogonal nearly-linear polarizations \cite{winkelnkemper_polarized_2007, winkelnkemper_polarized_2008, bardoux_polarized_2008}. This was illustrated in Fig.~\ref{fig:fss-theory} and observed in the A- and B-exciton PL spectra (Fig.~\ref{fig:AX_vs_BX}(a)). Moreover, the values of FSS and A- and B-exciton splitting could both be about 10~meV \cite{kindel_exciton_2010}. Hence unambiguous identification of FSS has been challenging. Here, when we change the QD from neutral to singly-charged, however, the FSS became suppressed while the A- and B-exciton splitting remained at about $20$~meV (Fig.~\ref{fig:AX_vs_BX}(b)). Hence the FSS from the A- and B-exciton splitting were clearly separated.

Finally, we distinguish the FSS or the A-B exciton splitting from the exciton-biexciton splitting. Orthogonal polarizations were measured between the fine structures and between the A- and B-excitons, while the same polarizations is expected for exciton and biexciton of the same branch \cite{amloy_polarization-resolved_2011}. In addition, as shown in Fig.~\ref{fig:AX_vs_BX}(c), the integrated intensities of both the $\X_\mathrm{A}^{0}$ and $\X_\mathrm{B}^{0}$ were approximately linearly dependent on the excitation density $P$ for $P<150$~W/cm$^2$, as expected of excitons and distinct from the $P^2$ dependence expected of bi-excitons.

In conclusion, we have demonstrated quantized charge tuning in site-controlled single InGaN/GaN QDs. Tunneling of an electron into an initially empty QD was controlled by an applied bias voltage and was manifested in discrete changes in the transition energy, static dipole moment and degeneracy of the corresponding excitonic state.  The ability to controllably charge an optically active QD with a single spin will allow the creation of Lambda-level system in the QD and thus coherent manipulation of the spin qubit, storage of quantum information in the spin-qubit and further quantum information processing.  The charge-tuning capability also allowed us to unambiguously identify and measure the fine structure splitting of the neutral exciton, simultaneously with the A- and B-exciton splitting and exciton-trion splitting. Such knowledge of the basic energy structure of the QDs provides experimental tests to previous theories regarding exciton charging and fine structures in III-Nitride QDs
\cite{winkelnkemper_polarized_2007, winkelnkemper_polarized_2008, bardoux_polarized_2008}, and is crucial for spintronics and spin-qubit quantum information processing, such as entanglement generation.

\appendix

\section{Photovoltage correction}
\label{appedix:photovoltage}

\begin{figure}
	\centering	\includegraphics[width=1\textwidth]{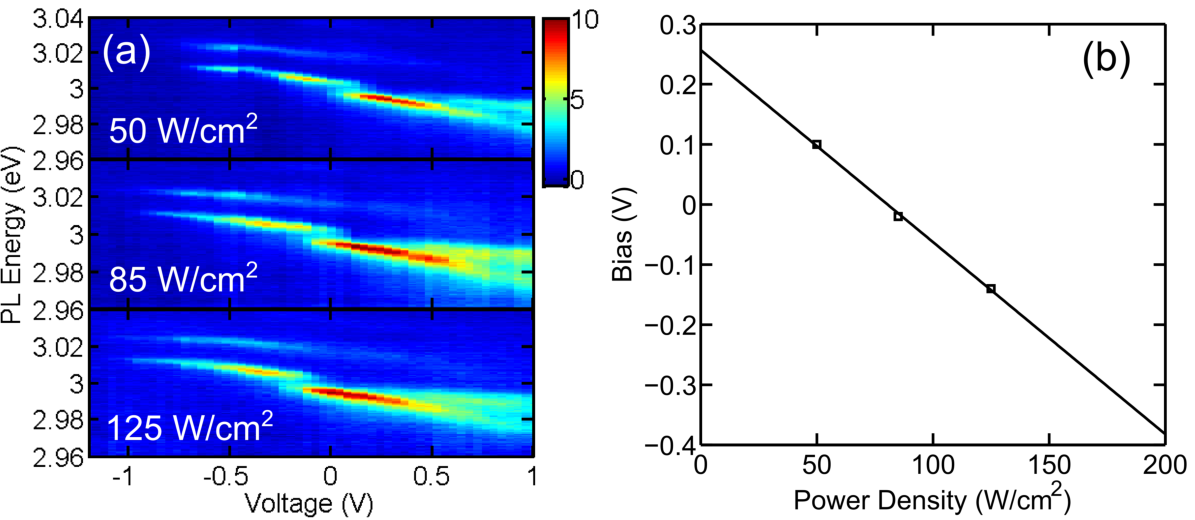}
	\caption{The photo-voltage effect. (a) PL energy $E$ vs. uncorrected voltage $V$ at excitation power densities $P = 50$, 85, and 125~W/cm$^2$, respectively. (d) The $\X_\mathrm{A}^0 \rightarrow \X_\mathrm{A}^{1-}$ transition voltage at the three $P$'s. Solid line is a linear fit.}
	\label{fig:photovoltage}
\end{figure}

Optical excitation generates free electrons and holes in the p-i-n junction region. Electrons and holes drift towards n-GaN and p-GaN, respectively, due to the built-in potential in the p-i-n junction. This photo-current results in a positive photo-voltage, i.e., p-contact has higher voltage than n-contact. As a result, the voltage-dependent PL energy ($E$-$V$) traces are shifted towards lower bias as the optical excitation power increases. This is shown in Fig.~\ref{fig:photovoltage}(a) in which the $E$-$V$ traces from the same QD are taken at different excitation power densities $P$. The same phenomenon has been observed in III-As based charged-tunneling diodes \cite{seidl_absorption_2005}. The external voltage $V_1$ required for the $\X_\mathrm{A}^0 \rightarrow \X_\mathrm{A}^{1-}$ transition in QD1 at different $P$'s are extracted in Fig.~\ref{fig:photovoltage}(b). The extrapolated $V_1$ at $P=0$ is the intrinsic bias voltage needed for $\X_\mathrm{A}^0 \rightarrow \X_\mathrm{A}^{1-}$ transition. Different QDs have different $V_1$-$P$ relation, so all $V_\mathrm{bias}$ values presented in the main text have been corrected according to their own $V_1$-$P$ data.

\section{Strain and band profile simulation}
\label{appedix:simulation}

\begin{figure}
	\centering	\includegraphics[width=1\textwidth]{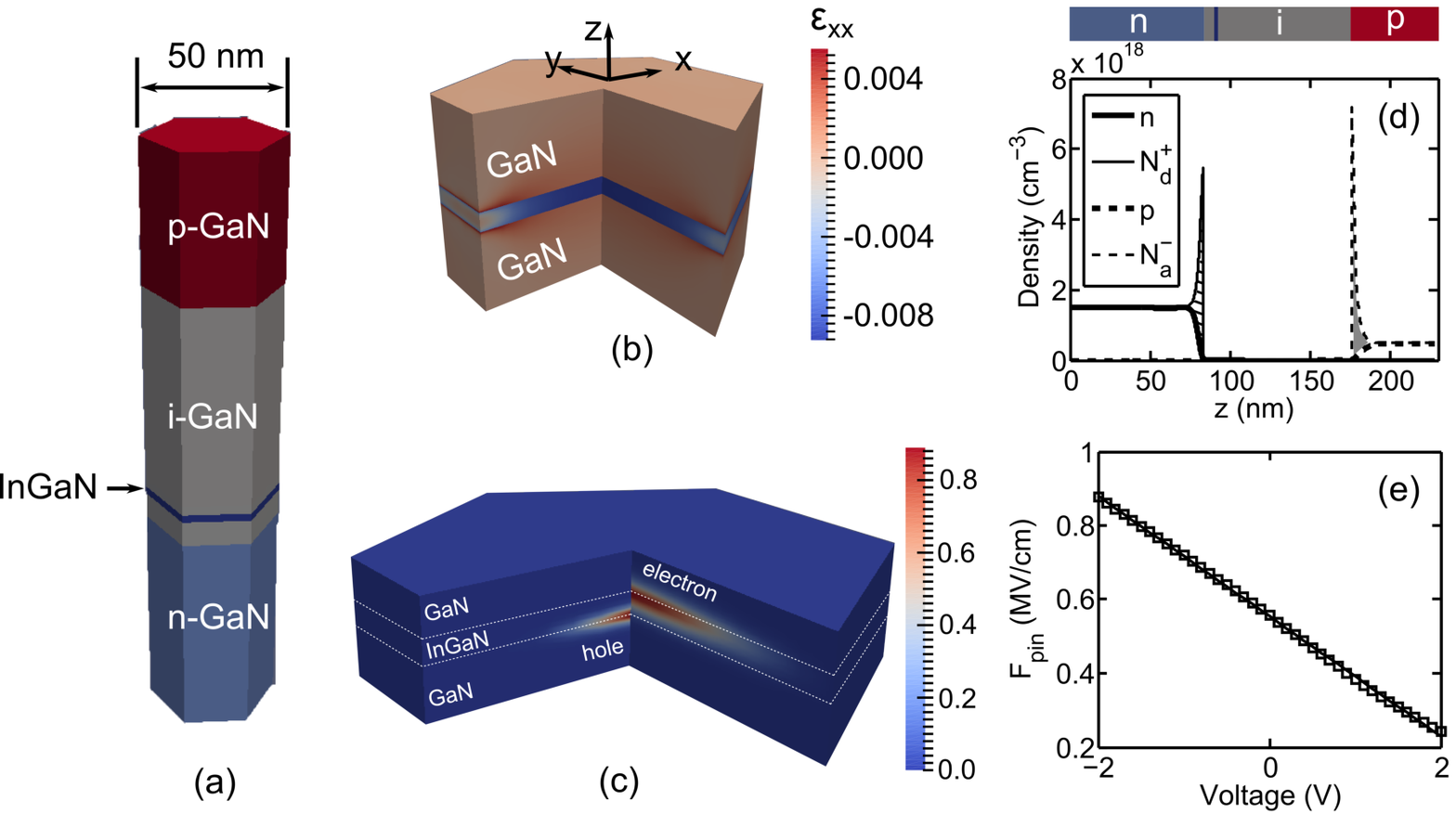}
	\caption{Strain, wavefunction and charge distribution simulation. (a) The structure of the nanowire. (b) $\epsilon_{xx}$ strain distribution near the InGaN nanodisk. (c) The cross section view of electron and hole ground state wavefunctions at bias voltage $V_\mathrm{bias} = 0$~V normalized by their own maximum. For better comparison, electron (hole) wavefunction is not shown in the xz (yz) cross section. (d) The $n$ (thick solid), $N_d^+$ (solid), $p$ (thick dashed), and $N_a^-$ (dashed) distribution along the z direction at the center of the nanowire. The top color bar shows the locations of different nanowire sections. The shadowed and gray area represent net positive and negative charge concentrations, respectively. (e) Calculated $F_\mathrm{pin}$ at different bias voltages $V_\mathrm{bias}$ (square) and the linear fitting line (solid).}
	\label{fig:simulation}
\end{figure}

The strain, band profiles and charge distribution of the device is calculated using a software package TiberCAD. The nanowire used in the simulation has a hexagonal cross section as shown in Fig.~\ref{fig:simulation}(a). The acceptor and donor densities in p- and n-GaN are $1\times10^{19}$~cm$^{-3}$ and $6\times 10^{18}$~cm$^{-3}$, respectively. They are obtained from the room temperature carrier concentrations $p = 5 \times 10^{17}$~cm$^{-3}$ and $n = 1.5 \times 10^{18}$~cm$^{-3}$ according to Hall measurement and assuming acceptor and donor levels being 150~meV \cite{orton_group_1999} and 15~meV \cite{gotz_activation_1996}, respectively. The hexagonal cross section is due to the selective wet-etching. We first calculate the strain distribution in the nanowire as shown in Fig.~\ref{fig:simulation}(b). The results suggest a strong strain relaxation at the edge of the InGaN nanodisk. The strain distribution is then used to calculate the bandgaps and polarization charges which are used in the subsequent calculation of band profiles by solving the Poisson and drift-diffusion equations. The conduction and valence band profiles along the central axis of the nanowire are shown in Fig.~\ref{fig:sample}. The internal polarization field $F_\mathrm{pol}$ at the center of the QD is obtained from the polarization charge density to be $1.6$~MV/cm pointing downward. Due to the non-uniform strain distribution in Fig.~\ref{fig:simulation}(b), the band profiles and electrical field close to the nanowire sidewalls are different from those at the center of the nanowire. However, since electron and hole wavefunctions are concentrated at the center of the InGaN nanodisk (Fig.~\ref{fig:simulation}(c)), the behavior of the carriers in the QD can be well understood using the results obtained at the center of the nanowire.

As mentioned in the main text, the p-i-n junction has a built-in electric field $F_\mathrm{pin}$ due to majority carrier diffusion. The electron (n), hole (p), ionized donor ($N_d^+$), and ionized acceptor ($N_a^-$) concentration at zero bias are shown in Fig.~\ref{fig:simulation}(d). The shadowed area represents the net positive charge concentration below the InGaN nanodisk, the gray area represents net negative charge concentration above the InGaN nanodisk. The $F_\mathrm{pin}$ can be calculated as following,
\begin{align}
F_\mathrm{pin} & = \frac{1}{2 \epsilon}(\sigma_+ + \sigma_-) \nonumber \\ 
& = \frac{1}{2 \epsilon} \left( \int N_d^+ - n \mathrm{d} z + \int N_a^- - p  \mathrm{d} z \right),
\end{align}
where $\epsilon$ is the permittivity of GaN. The calculated $F_\mathrm{pin}$ at various bias voltages $V_\mathrm{bias}$ is shown in Fig.~\ref{fig:simulation}(e). It fits very well with a linear function $F_\mathrm{pin} = F_\mathrm{pin0} - \eta V_\mathrm{bias}$, where $F_\mathrm{pin0} = 0.55$~MV/cm and $\eta = 0.16 \times 10^6$~/cm.

\begin{acknowledgments}
We acknowledge financial supports from the National Science Foundation (NSF) under Awards DMR 1409529 for work related to materials properties and device design and DMR 1120923 (MRSEC) for work related to the measurements. Part of the fabrication work was performed in the Lurie Nanofabrication Facility (LNF), which is part of the NSF NNIN network.
\end{acknowledgments}


\begin{thebibliography}{33}%
\makeatletter
\providecommand \@ifxundefined [1]{%
 \@ifx{#1\undefined}
}%
\providecommand \@ifnum [1]{%
 \ifnum #1\expandafter \@firstoftwo
 \else \expandafter \@secondoftwo
 \fi
}%
\providecommand \@ifx [1]{%
 \ifx #1\expandafter \@firstoftwo
 \else \expandafter \@secondoftwo
 \fi
}%
\providecommand \natexlab [1]{#1}%
\providecommand \enquote  [1]{``#1''}%
\providecommand \bibnamefont  [1]{#1}%
\providecommand \bibfnamefont [1]{#1}%
\providecommand \citenamefont [1]{#1}%
\providecommand \href@noop [0]{\@secondoftwo}%
\providecommand \href [0]{\begingroup \@sanitize@url \@href}%
\providecommand \@href[1]{\@@startlink{#1}\@@href}%
\providecommand \@@href[1]{\endgroup#1\@@endlink}%
\providecommand \@sanitize@url [0]{\catcode `\\12\catcode `\$12\catcode
  `\&12\catcode `\#12\catcode `\^12\catcode `\_12\catcode `\%12\relax}%
\providecommand \@@startlink[1]{}%
\providecommand \@@endlink[0]{}%
\providecommand \url  [0]{\begingroup\@sanitize@url \@url }%
\providecommand \@url [1]{\endgroup\@href {#1}{\urlprefix }}%
\providecommand \urlprefix  [0]{URL }%
\providecommand \Eprint [0]{\href }%
\providecommand \doibase [0]{http://dx.doi.org/}%
\providecommand \selectlanguage [0]{\@gobble}%
\providecommand \bibinfo  [0]{\@secondoftwo}%
\providecommand \bibfield  [0]{\@secondoftwo}%
\providecommand \translation [1]{[#1]}%
\providecommand \BibitemOpen [0]{}%
\providecommand \bibitemStop [0]{}%
\providecommand \bibitemNoStop [0]{.\EOS\space}%
\providecommand \EOS [0]{\spacefactor3000\relax}%
\providecommand \BibitemShut  [1]{\csname bibitem#1\endcsname}%
\let\auto@bib@innerbib\@empty
\bibitem [{\citenamefont {Warburton}(2013)}]{warburton_single_2013}%
  \BibitemOpen
  \bibfield  {author} {\bibinfo {author} {\bibfnamefont {R.~J.}\ \bibnamefont
  {Warburton}},\ }\href {http://dx.doi.org/10.1038/nmat3585} {\bibfield
  {journal} {\bibinfo  {journal} {Nature materials}\ }\textbf {\bibinfo
  {volume} {12}},\ \bibinfo {pages} {483} (\bibinfo {year} {2013})}\BibitemShut
  {NoStop}%
\bibitem [{\citenamefont {Kloeffel}\ and\ \citenamefont
  {Loss}(2013)}]{kloeffel_prospects_2013}%
  \BibitemOpen
  \bibfield  {author} {\bibinfo {author} {\bibfnamefont {C.}~\bibnamefont
  {Kloeffel}}\ and\ \bibinfo {author} {\bibfnamefont {D.}~\bibnamefont
  {Loss}},\ }\href {\doibase 10.1146/annurev-conmatphys-030212-184248}
  {\bibfield  {journal} {\bibinfo  {journal} {Annual Review of Condensed Matter
  Physics}\ }\textbf {\bibinfo {volume} {4}},\ \bibinfo {pages} {51} (\bibinfo
  {year} {2013})}\BibitemShut {NoStop}%
\bibitem [{\citenamefont {Warburton}\ \emph {et~al.}(2000)\citenamefont
  {Warburton}, \citenamefont {Sch\"{a}flein}, \citenamefont {Haft},
  \citenamefont {Bickel}, \citenamefont {Lorke}, \citenamefont {Karrai},
  \citenamefont {Garcia}, \citenamefont {Schoenfeld},\ and\ \citenamefont
  {Petroff}}]{warburton_optical_2000}%
  \BibitemOpen
  \bibfield  {author} {\bibinfo {author} {\bibfnamefont {R.~J.}\ \bibnamefont
  {Warburton}}, \bibinfo {author} {\bibfnamefont {C.}~\bibnamefont
  {Sch\"{a}flein}}, \bibinfo {author} {\bibfnamefont {D.}~\bibnamefont {Haft}},
  \bibinfo {author} {\bibfnamefont {F.}~\bibnamefont {Bickel}}, \bibinfo
  {author} {\bibfnamefont {A.}~\bibnamefont {Lorke}}, \bibinfo {author}
  {\bibfnamefont {K.}~\bibnamefont {Karrai}}, \bibinfo {author} {\bibfnamefont
  {J.~M.}\ \bibnamefont {Garcia}}, \bibinfo {author} {\bibfnamefont
  {W.}~\bibnamefont {Schoenfeld}}, \ and\ \bibinfo {author} {\bibfnamefont
  {P.~M.}\ \bibnamefont {Petroff}},\ }\href {\doibase 10.1038/35016030}
  {\bibfield  {journal} {\bibinfo  {journal} {Nature}\ }\textbf {\bibinfo
  {volume} {405}},\ \bibinfo {pages} {926} (\bibinfo {year}
  {2000})}\BibitemShut {NoStop}%
\bibitem [{\citenamefont {Gerardot}\ \emph {et~al.}(2008)\citenamefont
  {Gerardot}, \citenamefont {Brunner}, \citenamefont {Dalgarno}, \citenamefont
  {\"{O}hberg}, \citenamefont {Seidl}, \citenamefont {Kroner}, \citenamefont
  {Karrai}, \citenamefont {Stoltz}, \citenamefont {Petroff},\ and\
  \citenamefont {Warburton}}]{gerardot_optical_2005}%
  \BibitemOpen
  \bibfield  {author} {\bibinfo {author} {\bibfnamefont {B.~D.}\ \bibnamefont
  {Gerardot}}, \bibinfo {author} {\bibfnamefont {D.}~\bibnamefont {Brunner}},
  \bibinfo {author} {\bibfnamefont {P.~A.}\ \bibnamefont {Dalgarno}}, \bibinfo
  {author} {\bibfnamefont {P.}~\bibnamefont {\"{O}hberg}}, \bibinfo {author}
  {\bibfnamefont {S.}~\bibnamefont {Seidl}}, \bibinfo {author} {\bibfnamefont
  {M.}~\bibnamefont {Kroner}}, \bibinfo {author} {\bibfnamefont
  {K.}~\bibnamefont {Karrai}}, \bibinfo {author} {\bibfnamefont {N.~G.}\
  \bibnamefont {Stoltz}}, \bibinfo {author} {\bibfnamefont {P.~M.}\
  \bibnamefont {Petroff}}, \ and\ \bibinfo {author} {\bibfnamefont {R.~J.}\
  \bibnamefont {Warburton}},\ }\href
  {http://www.nature.com/doifinder/10.1038/nature06472} {\bibfield  {journal}
  {\bibinfo  {journal} {Nature}\ }\textbf {\bibinfo {volume} {451}},\ \bibinfo
  {pages} {441} (\bibinfo {year} {2008})}\BibitemShut {NoStop}%
\bibitem [{\citenamefont {Atat\"{u}re}\ \emph {et~al.}(2006)\citenamefont
  {Atat\"{u}re}, \citenamefont {Dreiser}, \citenamefont {Badolato},
  \citenamefont {H\"{o}gele}, \citenamefont {Karrai},\ and\ \citenamefont
  {Imamoglu}}]{atature_quantum-dot_2006}%
  \BibitemOpen
  \bibfield  {author} {\bibinfo {author} {\bibfnamefont {M.}~\bibnamefont
  {Atat\"{u}re}}, \bibinfo {author} {\bibfnamefont {J.}~\bibnamefont
  {Dreiser}}, \bibinfo {author} {\bibfnamefont {A.}~\bibnamefont {Badolato}},
  \bibinfo {author} {\bibfnamefont {A.}~\bibnamefont {H\"{o}gele}}, \bibinfo
  {author} {\bibfnamefont {K.}~\bibnamefont {Karrai}}, \ and\ \bibinfo {author}
  {\bibfnamefont {A.}~\bibnamefont {Imamoglu}},\ }\href
  {http://science.sciencemag.org/content/312/5773/551} {\bibfield  {journal}
  {\bibinfo  {journal} {Science}\ }\textbf {\bibinfo {volume} {312}},\ \bibinfo
  {pages} {551} (\bibinfo {year} {2006})}\BibitemShut {NoStop}%
\bibitem [{\citenamefont {Koppens}\ \emph {et~al.}(2006)\citenamefont
  {Koppens}, \citenamefont {Buizert}, \citenamefont {Tielrooij}, \citenamefont
  {Vink}, \citenamefont {Nowack}, \citenamefont {Meunier}, \citenamefont
  {Kouwenhoven},\ and\ \citenamefont {Vandersypen}}]{koppens_driven_2006}%
  \BibitemOpen
  \bibfield  {author} {\bibinfo {author} {\bibfnamefont {F.~H.~L.}\
  \bibnamefont {Koppens}}, \bibinfo {author} {\bibfnamefont {C.}~\bibnamefont
  {Buizert}}, \bibinfo {author} {\bibfnamefont {K.~J.}\ \bibnamefont
  {Tielrooij}}, \bibinfo {author} {\bibfnamefont {I.~T.}\ \bibnamefont {Vink}},
  \bibinfo {author} {\bibfnamefont {K.~C.}\ \bibnamefont {Nowack}}, \bibinfo
  {author} {\bibfnamefont {T.}~\bibnamefont {Meunier}}, \bibinfo {author}
  {\bibfnamefont {L.~P.}\ \bibnamefont {Kouwenhoven}}, \ and\ \bibinfo {author}
  {\bibfnamefont {L.~M.~K.}\ \bibnamefont {Vandersypen}},\ }\href {\doibase
  10.1038/nature05065} {\bibfield  {journal} {\bibinfo  {journal} {Nature}\
  }\textbf {\bibinfo {volume} {442}},\ \bibinfo {pages} {766} (\bibinfo {year}
  {2006})}\BibitemShut {NoStop}%
\bibitem [{\citenamefont {Xu}\ \emph {et~al.}(2007)\citenamefont {Xu},
  \citenamefont {Wu}, \citenamefont {Sun}, \citenamefont {Huang}, \citenamefont
  {Cheng}, \citenamefont {Steel}, \citenamefont {Bracker}, \citenamefont
  {Gammon}, \citenamefont {Emary},\ and\ \citenamefont {Sham}}]{xu_fast_2007}%
  \BibitemOpen
  \bibfield  {author} {\bibinfo {author} {\bibfnamefont {X.}~\bibnamefont
  {Xu}}, \bibinfo {author} {\bibfnamefont {Y.}~\bibnamefont {Wu}}, \bibinfo
  {author} {\bibfnamefont {B.}~\bibnamefont {Sun}}, \bibinfo {author}
  {\bibfnamefont {Q.}~\bibnamefont {Huang}}, \bibinfo {author} {\bibfnamefont
  {J.}~\bibnamefont {Cheng}}, \bibinfo {author} {\bibfnamefont {D.~G.}\
  \bibnamefont {Steel}}, \bibinfo {author} {\bibfnamefont {A.~S.}\ \bibnamefont
  {Bracker}}, \bibinfo {author} {\bibfnamefont {D.}~\bibnamefont {Gammon}},
  \bibinfo {author} {\bibfnamefont {C.}~\bibnamefont {Emary}}, \ and\ \bibinfo
  {author} {\bibfnamefont {L.~J.}\ \bibnamefont {Sham}},\ }\href
  {http://dx.doi.org/10.1103/PhysRevLett.99.097401} {\bibfield  {journal}
  {\bibinfo  {journal} {Physical Review Letters}\ }\textbf {\bibinfo {volume}
  {99}},\ \bibinfo {pages} {097401} (\bibinfo {year} {2007})}\BibitemShut
  {NoStop}%
\bibitem [{\citenamefont {Nowack}\ \emph {et~al.}(2007)\citenamefont {Nowack},
  \citenamefont {Koppens}, \citenamefont {Nazarov},\ and\ \citenamefont
  {Vandersypen}}]{nowack_coherent_2007}%
  \BibitemOpen
  \bibfield  {author} {\bibinfo {author} {\bibfnamefont {K.~C.}\ \bibnamefont
  {Nowack}}, \bibinfo {author} {\bibfnamefont {F.~H.~L.}\ \bibnamefont
  {Koppens}}, \bibinfo {author} {\bibfnamefont {Y.~V.}\ \bibnamefont
  {Nazarov}}, \ and\ \bibinfo {author} {\bibfnamefont {L.~M.~K.}\ \bibnamefont
  {Vandersypen}},\ }\href {http://www.sciencemag.org/content/318/5855/1430}
  {\bibfield  {journal} {\bibinfo  {journal} {Science}\ }\textbf {\bibinfo
  {volume} {318}},\ \bibinfo {pages} {1430} (\bibinfo {year}
  {2007})}\BibitemShut {NoStop}%
\bibitem [{\citenamefont {Press}\ \emph {et~al.}(2008)\citenamefont {Press},
  \citenamefont {Ladd}, \citenamefont {Zhang},\ and\ \citenamefont
  {Yamamoto}}]{press_complete_2008}%
  \BibitemOpen
  \bibfield  {author} {\bibinfo {author} {\bibfnamefont {D.}~\bibnamefont
  {Press}}, \bibinfo {author} {\bibfnamefont {T.~D.}\ \bibnamefont {Ladd}},
  \bibinfo {author} {\bibfnamefont {B.}~\bibnamefont {Zhang}}, \ and\ \bibinfo
  {author} {\bibfnamefont {Y.}~\bibnamefont {Yamamoto}},\ }\href {\doibase
  10.1038/nature07530} {\bibfield  {journal} {\bibinfo  {journal} {Nature}\
  }\textbf {\bibinfo {volume} {456}},\ \bibinfo {pages} {218} (\bibinfo {year}
  {2008})}\BibitemShut {NoStop}%
\bibitem [{\citenamefont {Kako}\ \emph {et~al.}(2006)\citenamefont {Kako},
  \citenamefont {Santori}, \citenamefont {Hoshino}, \citenamefont
  {G\"{o}tzinger}, \citenamefont {Yamamoto},\ and\ \citenamefont
  {Arakawa}}]{kako_gallium_2006}%
  \BibitemOpen
  \bibfield  {author} {\bibinfo {author} {\bibfnamefont {S.}~\bibnamefont
  {Kako}}, \bibinfo {author} {\bibfnamefont {C.}~\bibnamefont {Santori}},
  \bibinfo {author} {\bibfnamefont {K.}~\bibnamefont {Hoshino}}, \bibinfo
  {author} {\bibfnamefont {S.}~\bibnamefont {G\"{o}tzinger}}, \bibinfo {author}
  {\bibfnamefont {Y.}~\bibnamefont {Yamamoto}}, \ and\ \bibinfo {author}
  {\bibfnamefont {Y.}~\bibnamefont {Arakawa}},\ }\href {\doibase
  10.1038/nmat1763} {\bibfield  {journal} {\bibinfo  {journal} {Nature
  materials}\ }\textbf {\bibinfo {volume} {5}},\ \bibinfo {pages} {887}
  (\bibinfo {year} {2006})}\BibitemShut {NoStop}%
\bibitem [{\citenamefont {Holmes}\ \emph {et~al.}(2014)\citenamefont {Holmes},
  \citenamefont {Choi}, \citenamefont {Kako}, \citenamefont {Arita},\ and\
  \citenamefont {Arakawa}}]{holmes_room-temperature_2014}%
  \BibitemOpen
  \bibfield  {author} {\bibinfo {author} {\bibfnamefont {M.~J.}\ \bibnamefont
  {Holmes}}, \bibinfo {author} {\bibfnamefont {K.}~\bibnamefont {Choi}},
  \bibinfo {author} {\bibfnamefont {S.}~\bibnamefont {Kako}}, \bibinfo {author}
  {\bibfnamefont {M.}~\bibnamefont {Arita}}, \ and\ \bibinfo {author}
  {\bibfnamefont {Y.}~\bibnamefont {Arakawa}},\ }\href {\doibase
  10.1021/nl404400d} {\bibfield  {journal} {\bibinfo  {journal} {Nano letters}\
  }\textbf {\bibinfo {volume} {14}},\ \bibinfo {pages} {982} (\bibinfo {year}
  {2014})}\BibitemShut {NoStop}%
\bibitem [{\citenamefont {Hsu}\ \emph {et~al.}(2011)\citenamefont {Hsu},
  \citenamefont {Lundskog}, \citenamefont {Karlsson}, \citenamefont {Forsberg},
  \citenamefont {Janz\'{e}n},\ and\ \citenamefont {Holtz}}]{hsu_single_2011}%
  \BibitemOpen
  \bibfield  {author} {\bibinfo {author} {\bibfnamefont {C.-W.}\ \bibnamefont
  {Hsu}}, \bibinfo {author} {\bibfnamefont {A.}~\bibnamefont {Lundskog}},
  \bibinfo {author} {\bibfnamefont {K.~F.}\ \bibnamefont {Karlsson}}, \bibinfo
  {author} {\bibfnamefont {U.}~\bibnamefont {Forsberg}}, \bibinfo {author}
  {\bibfnamefont {E.}~\bibnamefont {Janz\'{e}n}}, \ and\ \bibinfo {author}
  {\bibfnamefont {P.~O.}\ \bibnamefont {Holtz}},\ }\href {\doibase
  10.1021/nl200810v} {\bibfield  {journal} {\bibinfo  {journal} {Nano letters}\
  }\textbf {\bibinfo {volume} {11}},\ \bibinfo {pages} {2415} (\bibinfo {year}
  {2011})}\BibitemShut {NoStop}%
\bibitem [{\citenamefont {Zhang}\ \emph {et~al.}(2013)\citenamefont {Zhang},
  \citenamefont {Teng}, \citenamefont {Hill}, \citenamefont {Lee},
  \citenamefont {Ku},\ and\ \citenamefont {Deng}}]{zhang_single_2013}%
  \BibitemOpen
  \bibfield  {author} {\bibinfo {author} {\bibfnamefont {L.}~\bibnamefont
  {Zhang}}, \bibinfo {author} {\bibfnamefont {C.-H.}\ \bibnamefont {Teng}},
  \bibinfo {author} {\bibfnamefont {T.~A.}\ \bibnamefont {Hill}}, \bibinfo
  {author} {\bibfnamefont {L.-K.}\ \bibnamefont {Lee}}, \bibinfo {author}
  {\bibfnamefont {P.-C.}\ \bibnamefont {Ku}}, \ and\ \bibinfo {author}
  {\bibfnamefont {H.}~\bibnamefont {Deng}},\ }\href {\doibase
  10.1063/1.4830000} {\bibfield  {journal} {\bibinfo  {journal} {Applied
  Physics Letters}\ }\textbf {\bibinfo {volume} {103}},\ \bibinfo {pages}
  {192114} (\bibinfo {year} {2013})}\BibitemShut {NoStop}%
\bibitem [{\citenamefont {Lundskog}\ \emph {et~al.}(2014)\citenamefont
  {Lundskog}, \citenamefont {Hsu}, \citenamefont {Karlsson}, \citenamefont
  {Amloy}, \citenamefont {Nilsson}, \citenamefont {Forsberg}, \citenamefont
  {{Olof Holtz}},\ and\ \citenamefont {Janz\'{e}n}}]{lundskog_direct_2014}%
  \BibitemOpen
  \bibfield  {author} {\bibinfo {author} {\bibfnamefont {A.}~\bibnamefont
  {Lundskog}}, \bibinfo {author} {\bibfnamefont {C.-W.}\ \bibnamefont {Hsu}},
  \bibinfo {author} {\bibfnamefont {K.~F.}\ \bibnamefont {Karlsson}}, \bibinfo
  {author} {\bibfnamefont {S.}~\bibnamefont {Amloy}}, \bibinfo {author}
  {\bibfnamefont {D.}~\bibnamefont {Nilsson}}, \bibinfo {author} {\bibfnamefont
  {U.}~\bibnamefont {Forsberg}}, \bibinfo {author} {\bibfnamefont
  {P.}~\bibnamefont {{Olof Holtz}}}, \ and\ \bibinfo {author} {\bibfnamefont
  {E.}~\bibnamefont {Janz\'{e}n}},\ }\href {\doibase 10.1038/lsa.2014.20}
  {\bibfield  {journal} {\bibinfo  {journal} {Light: Science \& Applications}\
  }\textbf {\bibinfo {volume} {3}},\ \bibinfo {pages} {e139} (\bibinfo {year}
  {2014})}\BibitemShut {NoStop}%
\bibitem [{\citenamefont {Teng}\ \emph {et~al.}(2015)\citenamefont {Teng},
  \citenamefont {Zhang}, \citenamefont {Hill}, \citenamefont {Demory},
  \citenamefont {Deng},\ and\ \citenamefont {Ku}}]{teng_elliptical_2015}%
  \BibitemOpen
  \bibfield  {author} {\bibinfo {author} {\bibfnamefont {C.-H.}\ \bibnamefont
  {Teng}}, \bibinfo {author} {\bibfnamefont {L.}~\bibnamefont {Zhang}},
  \bibinfo {author} {\bibfnamefont {T.~A.}\ \bibnamefont {Hill}}, \bibinfo
  {author} {\bibfnamefont {B.}~\bibnamefont {Demory}}, \bibinfo {author}
  {\bibfnamefont {H.}~\bibnamefont {Deng}}, \ and\ \bibinfo {author}
  {\bibfnamefont {P.-C.}\ \bibnamefont {Ku}},\ }\href
  {http://dx.doi.org/10.1063/1.4935463} {\bibfield  {journal} {\bibinfo
  {journal} {Applied Physics Letters}\ }\textbf {\bibinfo {volume} {107}},\
  \bibinfo {pages} {191105} (\bibinfo {year} {2015})}\BibitemShut {NoStop}%
\bibitem [{\citenamefont {Holmes}\ \emph {et~al.}(2013)\citenamefont {Holmes},
  \citenamefont {Kako}, \citenamefont {Choi}, \citenamefont {Podemski},
  \citenamefont {Arita},\ and\ \citenamefont
  {Arakawa}}]{holmes_measurement_2013}%
  \BibitemOpen
  \bibfield  {author} {\bibinfo {author} {\bibfnamefont {M.}~\bibnamefont
  {Holmes}}, \bibinfo {author} {\bibfnamefont {S.}~\bibnamefont {Kako}},
  \bibinfo {author} {\bibfnamefont {K.}~\bibnamefont {Choi}}, \bibinfo {author}
  {\bibfnamefont {P.}~\bibnamefont {Podemski}}, \bibinfo {author}
  {\bibfnamefont {M.}~\bibnamefont {Arita}}, \ and\ \bibinfo {author}
  {\bibfnamefont {Y.}~\bibnamefont {Arakawa}},\ }\href {\doibase
  10.1103/PhysRevLett.111.057401} {\bibfield  {journal} {\bibinfo  {journal}
  {Physical Review Letters}\ }\textbf {\bibinfo {volume} {111}},\ \bibinfo
  {pages} {057401} (\bibinfo {year} {2013})}\BibitemShut {NoStop}%
\bibitem [{\citenamefont {Songmuang}\ \emph {et~al.}(2010)\citenamefont
  {Songmuang}, \citenamefont {Katsaros}, \citenamefont {Monroy}, \citenamefont
  {Spathis}, \citenamefont {Bougerol}, \citenamefont {Mongillo},\ and\
  \citenamefont {De~Franceschi}}]{songmuang_quantum_2010}%
  \BibitemOpen
  \bibfield  {author} {\bibinfo {author} {\bibfnamefont {R.}~\bibnamefont
  {Songmuang}}, \bibinfo {author} {\bibfnamefont {G.}~\bibnamefont {Katsaros}},
  \bibinfo {author} {\bibfnamefont {E.}~\bibnamefont {Monroy}}, \bibinfo
  {author} {\bibfnamefont {P.}~\bibnamefont {Spathis}}, \bibinfo {author}
  {\bibfnamefont {C.}~\bibnamefont {Bougerol}}, \bibinfo {author}
  {\bibfnamefont {M.}~\bibnamefont {Mongillo}}, \ and\ \bibinfo {author}
  {\bibfnamefont {S.}~\bibnamefont {De~Franceschi}},\ }\href {\doibase
  10.1021/nl1017578} {\bibfield  {journal} {\bibinfo  {journal} {Nano Letters}\
  }\textbf {\bibinfo {volume} {10}},\ \bibinfo {pages} {3545} (\bibinfo {year}
  {2010})}\BibitemShut {NoStop}%
\bibitem [{\citenamefont {Lee}\ and\ \citenamefont
  {Ku}(2012)}]{lee_fabrication_2012}%
  \BibitemOpen
  \bibfield  {author} {\bibinfo {author} {\bibfnamefont {L.~K.}\ \bibnamefont
  {Lee}}\ and\ \bibinfo {author} {\bibfnamefont {P.-C.}\ \bibnamefont {Ku}},\
  }\href {\doibase 10.1002/pssc.201100428} {\bibfield  {journal} {\bibinfo
  {journal} {Physica Status Solidi (C)}\ }\textbf {\bibinfo {volume} {9}},\
  \bibinfo {pages} {609} (\bibinfo {year} {2012})}\BibitemShut {NoStop}%
\bibitem [{\citenamefont {Sacconi}\ \emph {et~al.}(2012)\citenamefont
  {Sacconi}, \citenamefont {Maur},\ and\ \citenamefont
  {Carlo}}]{sacconi_optoelectronic_2012}%
  \BibitemOpen
  \bibfield  {author} {\bibinfo {author} {\bibfnamefont {F.}~\bibnamefont
  {Sacconi}}, \bibinfo {author} {\bibfnamefont {M.~A.~D.}\ \bibnamefont
  {Maur}}, \ and\ \bibinfo {author} {\bibfnamefont {A.~D.}\ \bibnamefont
  {Carlo}},\ }\href {http://dx.doi.org/10.1109/TED.2012.2210897} {\bibfield
  {journal} {\bibinfo  {journal} {IEEE Transactions on Electron Devices}\
  }\textbf {\bibinfo {volume} {59}},\ \bibinfo {pages} {2979} (\bibinfo {year}
  {2012})}\BibitemShut {NoStop}%
\bibitem [{\citenamefont {Li}\ \emph {et~al.}(2011)\citenamefont {Li},
  \citenamefont {Westlake}, \citenamefont {Crawford}, \citenamefont {Lee},
  \citenamefont {Koleske}, \citenamefont {Figiel}, \citenamefont {Cross},
  \citenamefont {Fathololoumi}, \citenamefont {Mi},\ and\ \citenamefont
  {Wang}}]{li_optical_2011}%
  \BibitemOpen
  \bibfield  {author} {\bibinfo {author} {\bibfnamefont {Q.}~\bibnamefont
  {Li}}, \bibinfo {author} {\bibfnamefont {K.~R.}\ \bibnamefont {Westlake}},
  \bibinfo {author} {\bibfnamefont {M.~H.}\ \bibnamefont {Crawford}}, \bibinfo
  {author} {\bibfnamefont {S.~R.}\ \bibnamefont {Lee}}, \bibinfo {author}
  {\bibfnamefont {D.~D.}\ \bibnamefont {Koleske}}, \bibinfo {author}
  {\bibfnamefont {J.~J.}\ \bibnamefont {Figiel}}, \bibinfo {author}
  {\bibfnamefont {K.~C.}\ \bibnamefont {Cross}}, \bibinfo {author}
  {\bibfnamefont {S.}~\bibnamefont {Fathololoumi}}, \bibinfo {author}
  {\bibfnamefont {Z.}~\bibnamefont {Mi}}, \ and\ \bibinfo {author}
  {\bibfnamefont {G.~T.}\ \bibnamefont {Wang}},\ }\href
  {http://www.ncbi.nlm.nih.gov/pubmed/22273946} {\bibfield  {journal} {\bibinfo
   {journal} {Optics express}\ }\textbf {\bibinfo {volume} {19}},\ \bibinfo
  {pages} {25528} (\bibinfo {year} {2011})}\BibitemShut {NoStop}%
\bibitem [{\citenamefont {Seidl}\ \emph {et~al.}(2005)\citenamefont {Seidl},
  \citenamefont {Kroner}, \citenamefont {Dalgarno}, \citenamefont {H\"{o}gele},
  \citenamefont {Smith}, \citenamefont {Ediger}, \citenamefont {Gerardot},
  \citenamefont {Garcia}, \citenamefont {Petroff}, \citenamefont {Karrai},\
  and\ \citenamefont {Warburton}}]{seidl_absorption_2005}%
  \BibitemOpen
  \bibfield  {author} {\bibinfo {author} {\bibfnamefont {S.}~\bibnamefont
  {Seidl}}, \bibinfo {author} {\bibfnamefont {M.}~\bibnamefont {Kroner}},
  \bibinfo {author} {\bibfnamefont {P.~A.}\ \bibnamefont {Dalgarno}}, \bibinfo
  {author} {\bibfnamefont {A.}~\bibnamefont {H\"{o}gele}}, \bibinfo {author}
  {\bibfnamefont {J.~M.}\ \bibnamefont {Smith}}, \bibinfo {author}
  {\bibfnamefont {M.}~\bibnamefont {Ediger}}, \bibinfo {author} {\bibfnamefont
  {B.~D.}\ \bibnamefont {Gerardot}}, \bibinfo {author} {\bibfnamefont {J.~M.}\
  \bibnamefont {Garcia}}, \bibinfo {author} {\bibfnamefont {P.~M.}\
  \bibnamefont {Petroff}}, \bibinfo {author} {\bibfnamefont {K.}~\bibnamefont
  {Karrai}}, \ and\ \bibinfo {author} {\bibfnamefont {R.~J.}\ \bibnamefont
  {Warburton}},\ }\href {\doibase 10.1103/PhysRevB.72.195339} {\bibfield
  {journal} {\bibinfo  {journal} {Physical Review B}\ }\textbf {\bibinfo
  {volume} {72}},\ \bibinfo {pages} {195339} (\bibinfo {year}
  {2005})}\BibitemShut {NoStop}%
\bibitem [{\citenamefont {Baier}\ \emph {et~al.}(2001)\citenamefont {Baier},
  \citenamefont {Findeis}, \citenamefont {Zrenner}, \citenamefont {Bichler},\
  and\ \citenamefont {Abstreiter}}]{baier_optical_2001}%
  \BibitemOpen
  \bibfield  {author} {\bibinfo {author} {\bibfnamefont {M.}~\bibnamefont
  {Baier}}, \bibinfo {author} {\bibfnamefont {F.}~\bibnamefont {Findeis}},
  \bibinfo {author} {\bibfnamefont {A.}~\bibnamefont {Zrenner}}, \bibinfo
  {author} {\bibfnamefont {M.}~\bibnamefont {Bichler}}, \ and\ \bibinfo
  {author} {\bibfnamefont {G.}~\bibnamefont {Abstreiter}},\ }\href {\doibase
  10.1103/PhysRevB.64.195326} {\bibfield  {journal} {\bibinfo  {journal}
  {Physical Review B}\ }\textbf {\bibinfo {volume} {64}},\ \bibinfo {pages}
  {195326} (\bibinfo {year} {2001})}\BibitemShut {NoStop}%
\bibitem [{\citenamefont {Empedocles}\ and\ \citenamefont
  {Bawendi}(1997)}]{empedocles_quantum-confined_1997}%
  \BibitemOpen
  \bibfield  {author} {\bibinfo {author} {\bibfnamefont {S.~A.}\ \bibnamefont
  {Empedocles}}\ and\ \bibinfo {author} {\bibfnamefont {M.~G.}\ \bibnamefont
  {Bawendi}},\ }\href {\doibase 10.1126/science.278.5346.2114} {\bibfield
  {journal} {\bibinfo  {journal} {Science}\ }\textbf {\bibinfo {volume}
  {278}},\ \bibinfo {pages} {2114} (\bibinfo {year} {1997})}\BibitemShut
  {NoStop}%
\bibitem [{\citenamefont {Wang}\ \emph {et~al.}(2006)\citenamefont {Wang},
  \citenamefont {Shan}, \citenamefont {Islam}, \citenamefont {Herman},
  \citenamefont {Bonn},\ and\ \citenamefont {Heinz}}]{wang_exciton_2006}%
  \BibitemOpen
  \bibfield  {author} {\bibinfo {author} {\bibfnamefont {F.}~\bibnamefont
  {Wang}}, \bibinfo {author} {\bibfnamefont {J.}~\bibnamefont {Shan}}, \bibinfo
  {author} {\bibfnamefont {M.~A.}\ \bibnamefont {Islam}}, \bibinfo {author}
  {\bibfnamefont {I.~P.}\ \bibnamefont {Herman}}, \bibinfo {author}
  {\bibfnamefont {M.}~\bibnamefont {Bonn}}, \ and\ \bibinfo {author}
  {\bibfnamefont {T.~F.}\ \bibnamefont {Heinz}},\ }\href {\doibase
  10.1038/nmat1739} {\bibfield  {journal} {\bibinfo  {journal} {Nature
  materials}\ }\textbf {\bibinfo {volume} {5}},\ \bibinfo {pages} {861}
  (\bibinfo {year} {2006})}\BibitemShut {NoStop}%
\bibitem [{\citenamefont {Jarjour}\ \emph {et~al.}(2007)\citenamefont
  {Jarjour}, \citenamefont {Oliver}, \citenamefont {Tahraoui}, \citenamefont
  {Kappers}, \citenamefont {Humphreys},\ and\ \citenamefont
  {Taylor}}]{jarjour_control_2007}%
  \BibitemOpen
  \bibfield  {author} {\bibinfo {author} {\bibfnamefont {A.}~\bibnamefont
  {Jarjour}}, \bibinfo {author} {\bibfnamefont {R.}~\bibnamefont {Oliver}},
  \bibinfo {author} {\bibfnamefont {A.}~\bibnamefont {Tahraoui}}, \bibinfo
  {author} {\bibfnamefont {M.}~\bibnamefont {Kappers}}, \bibinfo {author}
  {\bibfnamefont {C.}~\bibnamefont {Humphreys}}, \ and\ \bibinfo {author}
  {\bibfnamefont {R.}~\bibnamefont {Taylor}},\ }\href {\doibase
  10.1103/PhysRevLett.99.197403} {\bibfield  {journal} {\bibinfo  {journal}
  {Physical Review Letters}\ }\textbf {\bibinfo {volume} {99}},\ \bibinfo
  {pages} {1} (\bibinfo {year} {2007})}\BibitemShut {NoStop}%
\bibitem [{\citenamefont {Amloy}\ \emph {et~al.}(2011)\citenamefont {Amloy},
  \citenamefont {Chen}, \citenamefont {Karlsson}, \citenamefont {Chen},
  \citenamefont {Hsu}, \citenamefont {Hsiao}, \citenamefont {Chen},\ and\
  \citenamefont {Holtz}}]{amloy_polarization-resolved_2011}%
  \BibitemOpen
  \bibfield  {author} {\bibinfo {author} {\bibfnamefont {S.}~\bibnamefont
  {Amloy}}, \bibinfo {author} {\bibfnamefont {Y.~T.}\ \bibnamefont {Chen}},
  \bibinfo {author} {\bibfnamefont {K.~F.}\ \bibnamefont {Karlsson}}, \bibinfo
  {author} {\bibfnamefont {K.~H.}\ \bibnamefont {Chen}}, \bibinfo {author}
  {\bibfnamefont {H.~C.}\ \bibnamefont {Hsu}}, \bibinfo {author} {\bibfnamefont
  {C.~L.}\ \bibnamefont {Hsiao}}, \bibinfo {author} {\bibfnamefont {L.~C.}\
  \bibnamefont {Chen}}, \ and\ \bibinfo {author} {\bibfnamefont {P.~O.}\
  \bibnamefont {Holtz}},\ }\href {\doibase 10.1103/PhysRevB.83.201307}
  {\bibfield  {journal} {\bibinfo  {journal} {Physical Review B}\ }\textbf
  {\bibinfo {volume} {83}},\ \bibinfo {pages} {201307} (\bibinfo {year}
  {2011})}\BibitemShut {NoStop}%
\bibitem [{\citenamefont {Chuang}\ and\ \citenamefont
  {Chang}(1996)}]{chuang_kp_1996}%
  \BibitemOpen
  \bibfield  {author} {\bibinfo {author} {\bibfnamefont {S.~L.}\ \bibnamefont
  {Chuang}}\ and\ \bibinfo {author} {\bibfnamefont {C.~S.}\ \bibnamefont
  {Chang}},\ }\href {\doibase 10.1103/PhysRevB.54.2491} {\bibfield  {journal}
  {\bibinfo  {journal} {Physical Review B}\ }\textbf {\bibinfo {volume} {54}},\
  \bibinfo {pages} {2491} (\bibinfo {year} {1996})}\BibitemShut {NoStop}%
\bibitem [{\citenamefont {Winkelnkemper}\ \emph {et~al.}(2007)\citenamefont
  {Winkelnkemper}, \citenamefont {Seguin}, \citenamefont {Rodt}, \citenamefont
  {Schliwa}, \citenamefont {Rei{\ss}mann}, \citenamefont {Strittmatter},
  \citenamefont {Hoffmann},\ and\ \citenamefont
  {Bimberg}}]{winkelnkemper_polarized_2007}%
  \BibitemOpen
  \bibfield  {author} {\bibinfo {author} {\bibfnamefont {M.}~\bibnamefont
  {Winkelnkemper}}, \bibinfo {author} {\bibfnamefont {R.}~\bibnamefont
  {Seguin}}, \bibinfo {author} {\bibfnamefont {S.}~\bibnamefont {Rodt}},
  \bibinfo {author} {\bibfnamefont {A.}~\bibnamefont {Schliwa}}, \bibinfo
  {author} {\bibfnamefont {L.}~\bibnamefont {Rei{\ss}mann}}, \bibinfo {author}
  {\bibfnamefont {A.}~\bibnamefont {Strittmatter}}, \bibinfo {author}
  {\bibfnamefont {A.}~\bibnamefont {Hoffmann}}, \ and\ \bibinfo {author}
  {\bibfnamefont {D.}~\bibnamefont {Bimberg}},\ }\href {\doibase
  10.1063/1.2743893} {\bibfield  {journal} {\bibinfo  {journal} {Journal of
  Applied Physics}\ }\textbf {\bibinfo {volume} {101}},\ \bibinfo {pages}
  {113708} (\bibinfo {year} {2007})}\BibitemShut {NoStop}%
\bibitem [{\citenamefont {Winkelnkemper}\ \emph {et~al.}(2008)\citenamefont
  {Winkelnkemper}, \citenamefont {Seguin}, \citenamefont {Rodt}, \citenamefont
  {Schliwa}, \citenamefont {Reismann}, \citenamefont {Strittmatter},
  \citenamefont {Hoffmann},\ and\ \citenamefont
  {Bimberg}}]{winkelnkemper_polarized_2008}%
  \BibitemOpen
  \bibfield  {author} {\bibinfo {author} {\bibfnamefont {M.}~\bibnamefont
  {Winkelnkemper}}, \bibinfo {author} {\bibfnamefont {R.}~\bibnamefont
  {Seguin}}, \bibinfo {author} {\bibfnamefont {S.}~\bibnamefont {Rodt}},
  \bibinfo {author} {\bibfnamefont {A.}~\bibnamefont {Schliwa}}, \bibinfo
  {author} {\bibfnamefont {L.}~\bibnamefont {Reismann}}, \bibinfo {author}
  {\bibfnamefont {A.}~\bibnamefont {Strittmatter}}, \bibinfo {author}
  {\bibfnamefont {A.}~\bibnamefont {Hoffmann}}, \ and\ \bibinfo {author}
  {\bibfnamefont {D.}~\bibnamefont {Bimberg}},\ }\href {\doibase
  10.1016/j.physe.2007.11.033} {\bibfield  {journal} {\bibinfo  {journal}
  {Physica E: Low-dimensional Systems and Nanostructures}\ }\textbf {\bibinfo
  {volume} {40}},\ \bibinfo {pages} {2217} (\bibinfo {year}
  {2008})}\BibitemShut {NoStop}%
\bibitem [{\citenamefont {Bardoux}\ \emph {et~al.}(2008)\citenamefont
  {Bardoux}, \citenamefont {Guillet}, \citenamefont {Gil}, \citenamefont
  {Lefebvre}, \citenamefont {Bretagnon}, \citenamefont {Taliercio},
  \citenamefont {Rousset},\ and\ \citenamefont
  {Semond}}]{bardoux_polarized_2008}%
  \BibitemOpen
  \bibfield  {author} {\bibinfo {author} {\bibfnamefont {R.}~\bibnamefont
  {Bardoux}}, \bibinfo {author} {\bibfnamefont {T.}~\bibnamefont {Guillet}},
  \bibinfo {author} {\bibfnamefont {B.}~\bibnamefont {Gil}}, \bibinfo {author}
  {\bibfnamefont {P.}~\bibnamefont {Lefebvre}}, \bibinfo {author}
  {\bibfnamefont {T.}~\bibnamefont {Bretagnon}}, \bibinfo {author}
  {\bibfnamefont {T.}~\bibnamefont {Taliercio}}, \bibinfo {author}
  {\bibfnamefont {S.}~\bibnamefont {Rousset}}, \ and\ \bibinfo {author}
  {\bibfnamefont {F.}~\bibnamefont {Semond}},\ }\href {\doibase
  10.1103/PhysRevB.77.235315} {\bibfield  {journal} {\bibinfo  {journal}
  {Physical Review B}\ }\textbf {\bibinfo {volume} {77}},\ \bibinfo {pages}
  {235315} (\bibinfo {year} {2008})}\BibitemShut {NoStop}%
\bibitem [{\citenamefont {Kindel}\ \emph {et~al.}(2010)\citenamefont {Kindel},
  \citenamefont {Kako}, \citenamefont {Kawano}, \citenamefont {Oishi},
  \citenamefont {Arakawa}, \citenamefont {H\"{o}nig}, \citenamefont
  {Winkelnkemper}, \citenamefont {Schliwa}, \citenamefont {Hoffmann},\ and\
  \citenamefont {Bimberg}}]{kindel_exciton_2010}%
  \BibitemOpen
  \bibfield  {author} {\bibinfo {author} {\bibfnamefont {C.}~\bibnamefont
  {Kindel}}, \bibinfo {author} {\bibfnamefont {S.}~\bibnamefont {Kako}},
  \bibinfo {author} {\bibfnamefont {T.}~\bibnamefont {Kawano}}, \bibinfo
  {author} {\bibfnamefont {H.}~\bibnamefont {Oishi}}, \bibinfo {author}
  {\bibfnamefont {Y.}~\bibnamefont {Arakawa}}, \bibinfo {author} {\bibfnamefont
  {G.}~\bibnamefont {H\"{o}nig}}, \bibinfo {author} {\bibfnamefont
  {M.}~\bibnamefont {Winkelnkemper}}, \bibinfo {author} {\bibfnamefont
  {A.}~\bibnamefont {Schliwa}}, \bibinfo {author} {\bibfnamefont
  {A.}~\bibnamefont {Hoffmann}}, \ and\ \bibinfo {author} {\bibfnamefont
  {D.}~\bibnamefont {Bimberg}},\ }\href {\doibase 10.1103/PhysRevB.81.241309}
  {\bibfield  {journal} {\bibinfo  {journal} {Physical Review B}\ }\textbf
  {\bibinfo {volume} {81}},\ \bibinfo {pages} {241309} (\bibinfo {year}
  {2010})}\BibitemShut {NoStop}%
\bibitem [{\citenamefont {Orton}\ and\ \citenamefont
  {Foxon}(1999)}]{orton_group_1999}%
  \BibitemOpen
  \bibfield  {author} {\bibinfo {author} {\bibfnamefont {J.~W.}\ \bibnamefont
  {Orton}}\ and\ \bibinfo {author} {\bibfnamefont {C.~T.}\ \bibnamefont
  {Foxon}},\ }\href {\doibase 10.1088/0034-4885/61/1/001} {\bibfield  {journal}
  {\bibinfo  {journal} {Reports on Progress in Physics}\ }\textbf {\bibinfo
  {volume} {61}},\ \bibinfo {pages} {1} (\bibinfo {year} {1999})}\BibitemShut
  {NoStop}%
\bibitem [{\citenamefont {Götz}\ \emph {et~al.}(1996)\citenamefont {Götz},
  \citenamefont {Johnson}, \citenamefont {Chen}, \citenamefont {Liu},
  \citenamefont {Kuo},\ and\ \citenamefont {Imler}}]{gotz_activation_1996}%
  \BibitemOpen
  \bibfield  {author} {\bibinfo {author} {\bibfnamefont {W.}~\bibnamefont
  {Götz}}, \bibinfo {author} {\bibfnamefont {N.~M.}\ \bibnamefont {Johnson}},
  \bibinfo {author} {\bibfnamefont {C.}~\bibnamefont {Chen}}, \bibinfo {author}
  {\bibfnamefont {H.}~\bibnamefont {Liu}}, \bibinfo {author} {\bibfnamefont
  {C.}~\bibnamefont {Kuo}}, \ and\ \bibinfo {author} {\bibfnamefont
  {W.}~\bibnamefont {Imler}},\ }\href {\doibase 10.1063/1.115805} {\bibfield
  {journal} {\bibinfo  {journal} {Applied Physics Letters}\ }\textbf {\bibinfo
  {volume} {68}},\ \bibinfo {pages} {3144} (\bibinfo {year}
  {1996})}\BibitemShut {NoStop}%
\end{thebibliography}
%

\end{document}